\documentclass[prl,floatfix,twocolumn]{revtex4}
\usepackage{amsfonts}
\usepackage{amsmath}
\usepackage{amssymb}
\usepackage{dcolumn}
\usepackage{bm}
\usepackage{epsfig}
\usepackage{graphicx,graphics}
\usepackage{wrapfig}
\usepackage{dcolumn}
\usepackage{fancybox} % sorry; bbm stuck on my latex
\usepackage{euscript}
\usepackage[german,english]{babel}
\usepackage{psfrag}
\usepackage{color}
\begin{document}

\title{Quantum computation in correlation space and extremal entanglement}
\author{J.-M. Cai$^{1,2}$, W. D\"ur$^{1,2}$, M. Van den Nest$^3$, A. Miyake$^{1,2,4}$, and H. J. Briegel$^{1,2}$}
\affiliation{$^1$Institut f\"ur Quantenoptik und Quanteninformation der \"Osterreichischen Akademie der Wissenschaften, Innsbruck, Austria\\
$^2$Institut f{\"u}r Theoretische Physik, Universit{\"a}t Innsbruck, Technikerstra{\ss}e 25, A-6020 Innsbruck, Austria\\
$^3$Max-Planck-Institut f\"ur Quantenoptik, Hans-Kopfermann-Str. 1, D-85748 Garching, Germany\\
$^4$Perimeter Institute for Theoretical Physics, 31 Caroline St. N.,
Waterloo ON, N2L 2Y5, Canada}

\begin{abstract}
Recently, a framework was established to systematically construct
novel universal resource states for measurement-based quantum
computation using techniques involving finitely correlated states.
With these methods, universal states were found which are in certain
ways much less entangled than the original cluster state model, and
it was hence believed that with this approach many of the extremal
entanglement features of the cluster states could be relaxed. The
new resources were constructed as ``computationally universal''
states---i.e. they allow one to efficiently reproduce the classical
output of each quantum computation---whereas the cluster states are
universal in a stronger sense since they are ``universal state
preparators''.  Here we show that the new resources are universal
state preparators after all, and must therefore exhibit a whole
class of extremal entanglement features, similar to the cluster
states.
\end{abstract}

\maketitle

{\bf Introduction.---} The paradigm of the one-way quantum computer
\cite{Ra01} established that there exist certain states of large
scale quantum systems---the cluster states \cite{Br01}---with the
remarkable feature that universal quantum computation can be
achieved by merely performing single-qubit measurements on these
states. This new computational model has opened a novel approach
towards the possible experimental realization of quantum
computation, but also provides a different perspective to study
fundamental questions regarding the power and limitations of quantum
computation \cite{Br09}. The properties of the cluster-state scheme
have been intensively studied, and it was found that these states
exhibit several rather singular features: e.g., they are highly
entangled \cite{Br01, Br09, He06, Va06a}, have vanishing two-point
correlation functions \cite{He06} and cannot occur as non-degenerate
ground states of two-local spin-$1/2$ Hamiltonians \cite{Ni05}.
These properties are sometimes viewed as a drawback of the scheme,
since they indicate that it may be not be easy to find cluster
states as states of naturally occurring physical systems. This has
raised the question whether there might exist other universal
resource states for which some of these extremal features could be
relaxed and which are easier to prepare experimentally (see also
\cite{Ba06, Br08, Chen08}). More fundamentally, this connects to the
important problem of understanding which requirements must be met by
every universal resource state.

Recently \cite{Gr06, Gr07, Gr08}, a framework was proposed to
construct novel universal resource states. The approach, sometimes
called ``quantum computation in correlation space''
($\mathcal{QC}_{cs}$), uses techniques known from the study of
finitely correlated states \cite{Fa92} and projected-entangled-pairs
states \cite{Ve04,Ve07}. Interestingly, with this approach universal
states have been constructed in which several of the above extremal
features are indeed no longer present. For example, universal states
have been found which are locally arbitrarily pure and which have
non-zero two-point correlations \cite{Gr06, Gr07}.

These results raised the impression that the $\mathcal{QC}_{cs}$
universal states are fundamentally different from the cluster
states, and that extremal entanglement features are generally not
necessary to achieve universality. The aim of this construction was
to obtain states that are universal in a strictly weaker sense than
the cluster states. Upon closer inspection \cite{Va06a} one finds
that the cluster states are ``universal state preparators'': they
allow to (efficiently) reproduce the \emph{quantum output} of every
quantum computation. The $\mathcal{QC}_{cs}$ universal states,
instead, are ``computationally universal'': with the help of such a
universal state, it is possible to efficiently reproduce the
\emph{classical output} of every quantum computation. As shown in
\cite{Va06a}, every universal state preparator must exhibit a whole
class of extremal entanglement features. Whether or not the
universal $\mathcal{QC}_{cs}$ states are universal state preparators
as well, was initially unclear. It seemed as if, the stringent
entanglement criteria for universal state preparators did not apply
to the $\mathcal{QC}_{cs}$ universal states.

As the main result of this letter, we show that, under rather general
conditions, $\mathcal{QC}_{cs}$ universal states are universal state
preparators after all---and therefore have to meet all extremal
entanglement requirements for universal state preparators as derived
in \cite{Va06a}. Our results apply in particular to
the general construction of universal $\mathcal{QC}_{cs}$ resources
presented in \cite{Gr08}.

\begin{figure}[tbh]
\epsfig{file=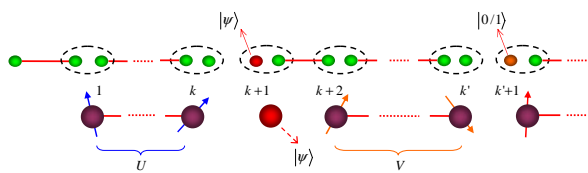,width=8cm} \caption{(Color online)
$\mathcal{QC}_{cs}$ in a 1D wire, where the single-qubit output
state $|\psi\rangle$ is localized at the physical site $k+1$. (i)
Local measurements (blue arrows) on the physical sites $1$ to $k$
implement the desired unitary rotation $U$ on the correlation
system, the output quantum state $ \vert \psi\rangle$ is in the red
virtual qubit. (ii) Local measurements (brown arrows) on the
physical sites $k+2$ to $k'$ implement a specific unitary
transformation $V= \vert +\rangle\langle \varphi_{0} \vert + \vert
-\rangle\langle \varphi_{1} \vert $, and a measurement in the basis
$ \vert m_{0}\rangle,  \vert m_{1}\rangle$ at site $k'+1$ allows one
to localize the quantum information at site $k+1$.}
\label{specificquwire}
\end{figure}

{\bf Quantum computation in correlation space.---}
$\mathcal{QC}_{cs}$ has been introduced as a scheme for
measurement-based quantum computation that enables one to
systematically construct universal resource states \cite{Gr06, Gr07,
Gr08}. A resource state for $\mathcal{QC}_{cs}$ is described by a
computational tensor network or, equivalently, by a
projected-entangled-pairs state or valence bond state
\cite{Ve07,Ve04} with certain boundary conditions (see
Fig.~\ref{specificquwire}). Quantum information is processed in a
virtual system, the ``correlation space'', that consists of
entangled pairs of $D$-dimensional systems arranged on a lattice.
Each physical particle is associated with $t$ of these virtual
systems by means of a projection (or tensor), mapping the
$D^t$-dimensional ``virtual'' space to the $d$-dimensional
``physical'' space. It can be shown that single-qubit projective
measurements on the physical particles induce operations in the
virtual system, and, for properly chosen resource states, in this
way every quantum state $|\psi_{\mbox{\scriptsize{out}}}\rangle = U
\vert 0\rangle^{\otimes n}$ (where $U$ is any poly-size unitary
operation) can be (efficiently) prepared in the virtual system.
Furthermore, the quantum information stored within the correlation
system can be read out by local projective measurements on the
physical sites as well. Hence, it is possible to efficiently
reproduce the measurement statistics of every quantum computation,
even though the output state
$|\psi_{\mbox{\scriptsize{out}}}\rangle$ is never prepared in the
physical Hilbert space but rather in the abstract correlation space.
Therefore, the $\mathcal{QC}_{cs}$ resource states are naturally
computationally universal, but at first sight do not appear to be
universal state preparators. Here, however,we provide an explicit
protocol to prepare $|\psi_{\rm out}\rangle$ in the physical Hilbert
space for a large class of $\mathcal{QC}_{cs}$ universal states.
Following \cite{Va06a}, we thereby consider the slightly more
general class of local measurements, including positive
operator-valued measures (POVM), and classical communications, and
show that these resources are universal state preparators after all.

General universal $\mathcal{QC}_{cs}$ resources can be constructed
by connecting 1D quantum computational ``wires'', which are matrix
product states that allow processing of one logical qubit, into
universal 2D quantum ``webs'' \cite{Gr08}. Here, we will follow a
similar line of argument as in \cite{Gr08}, by first considering the
1D quantum wires and then 2D webs. We will show that quantum
information stored in the correlation system can be localized at
physical sites by a simple procedure. To simplify the discussion, we
will consider $D=d=2$ in the following; however, our results are not
restricted to this case.

%\begin{figure}[tbh]
%\epsfig{file=Fig1.eps,width=8cm} \caption{(Color online)
%$\mathcal{QC}_{cs}$ in a 1D wire, where the single-qubit output
%state $|\psi\rangle$ is localized at the physical site
%$k+1$.}\label{specificquwire}
%\end{figure}

{\bf Quantum computation on 1D resources.---} We start from 1D
quantum wires described by matrix product states (MPS)
\cite{Gr06,Gr07}
\begin{equation}
\label{phiL} \Phi\left( \vert L\rangle \right)_{1}^{ N}
=\sum\limits_{s_{i}=0, 1}\langle R \vert A[{s_{n}}]\cdots A[{s_{1}}]
\vert L\rangle\  \vert s_{1}\cdots s_{n}\rangle
\end{equation}
where $1$ and $N$ denote the starting and ending site, $ \vert
L\rangle$ and $ \vert R\rangle$ represent left and right boundary
conditions, and we assume in the following that $ \vert R\rangle$
is fixed. The matrices $A[s_j]$ are determined by the projections
from the virtual system to the physical system, see
Fig.~\ref{specificquwire}.

The state $\Phi\left( \vert L\rangle \right)_{1}^{ N}$ is called a
universal quantum wire if, by performing suitable single-qubit
measurements on, say, the first $k$ sites, it is possible to
efficiently prepare $\Phi\left( \vert \psi\rangle \right)_{k+1}^{N}$
(cf. Eq. (\ref{phiL})) on the unmeasured qubits, for every one-qubit
state $|\psi\rangle$. We will show that, for each universal quantum
wire, it is possible to post-process the state $\Phi\left( \vert
\psi\rangle \right)_{k+1}^{N}$ by local measurements, in such a way
that the state $|\psi\rangle$ is efficiently prepared, i.e.
localized, in one of the physical sites of the chain.

In order to demonstrate our basic idea, we first consider a simple
instance of a universal quantum wire, which nevertheless covers many
of the resources in \cite{Gr06,Gr07}: we assume that there exists a
local basis $\{ \vert m_{s}\rangle\}$ such that
\begin{equation}
A[{m_{0}}]= \vert \varphi_{0}\rangle \langle 0 \vert ,  \quad
A[{m_{1}}]= \vert \varphi_{1}\rangle \langle 1 \vert
\label{tworankone}
\end{equation}
where $A[{m_{s}}]=\langle m_{s} \vert 0\rangle A[0]+\langle m_{s}
\vert 1\rangle A[1]$ with $\langle \varphi _{i} \vert \varphi
_{j}\rangle=\delta_{ij}$. For such a state, by performing local
measurements on the first $k$ physical sites, one can indeed
efficiently prepare $\Phi\left( \vert \psi\rangle
\right)_{k+1}^{N}$, for every  $\vert \psi \rangle =\lambda_{0}
\vert 0\rangle +\lambda _{1} \vert 1\rangle$ \cite{Gr06,Gr07}.
%\end{equation}
The output quantum state $ \vert \psi \rangle$ is associated with
the virtual qubit corresponding to the physical site $k+1$ (i.e. the
red virtual qubit in Fig.\ref{specificquwire}).  Our goal is now to
``localize'' the state $ \vert \psi\rangle$ in one of the physical
sites by performing further measurements. This will proceed in two
steps.

(i) We rewrite the state  $\Phi\left( \vert \psi\rangle
\right)_{k+1}^{ N}$ in terms of the basis $\{ \vert m_{s}\rangle\}$
at site $k+1$:
\begin{equation}
\Phi( \vert \psi\rangle)_{k+1}^N=\sum_{s=0,1}\lambda_s \vert
m_{s}\rangle_{k+1} \Phi( \vert \varphi_s\rangle)_{k+2}^N
\label{phikplusone}
\end{equation}
Although this state can already be viewed as an encoded version of the
output state $ \vert \psi\rangle$ with the block encoding $ \vert
s_L\rangle \equiv  \vert m_{s}\rangle_{k+1} \Phi( \vert
\varphi_s\rangle)_{k+2}^N$, the quantum output is carried by all the
unmeasured physical sites.

(ii) By  performing suitable local measurements on the physical
sites from $k+2$ to, say, $k'$ (see Fig.~\ref{specificquwire}) one
can implement the unitary single qubit rotation $V$ which maps $
\vert \varphi_{0}\rangle \rightarrow  \vert +\rangle$ and $ \vert
\varphi_{1}\rangle \rightarrow  \vert -\rangle$, where $ \vert
\pm\rangle$ denote the Pauli $X$ eigenstates. This is possible by
the very assumption that the MPS defined by $\{A[0], A[1]\}$ is a
universal computational wire---i.e. the two states $\Phi( \vert
\varphi_s\rangle)_{k+2}^N$ can further be used to implement
arbitrary single-qubit rotations in the correlation space, where in
fact the required measurement pattern is the same in both cases. In
this way,  the state (\ref{phikplusone}) is mapped to
\begin{equation}
\lambda_0 \vert m_{0}\rangle_{k+1} \Phi( \vert +\rangle)_{k'+1}^N +
\lambda_1 \vert m_{1}\rangle_{k+1} \Phi( \vert -\rangle)_{k'+1}^N.
\end{equation}
Expanding site $k'+1$ in the basis $\{ \vert m_s\rangle\}$, it
can be readily shown that the resulting state is equal to
\begin{equation}
\sum_{s=0,1} Z_m^s \vert \psi_m\rangle_{k+1}  \vert
m_s\rangle_{k'+1} \Phi( \vert \varphi_s\rangle)_{k'+2}^N,
\end{equation}
where $ \vert \psi_m\rangle =\lambda _{0} \vert  m_{0}\rangle
+\lambda _{1} \vert  m_{1}\rangle$ and $Z_m= \vert
m_{0}\rangle\langle m_{0} \vert  - \vert  m_{1}\rangle\langle m_{1}
\vert $. By measuring the physical site $k'+1$ in the basis $\{
\vert m_{s}\rangle\}$, we obtain the (encoded) output quantum state
$ \vert \psi_m\rangle$ at site $k+1$, with possible local Pauli
correction $Z_m$, which is now decoupled from the remaining wire.
That is, the desired output quantum state is generated at the
physical site $k+1$ in the basis $\{ \vert m_0\rangle, \vert
m_1\rangle\}$.

\begin{figure}[tbh]
\epsfig{file=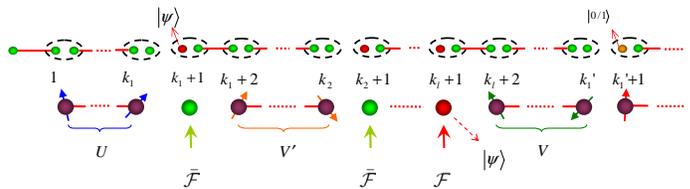,width=9cm} \caption{ (Color online) Generation
of arbitrary single-qubit output states from general quantum wires
by Eq.~(\ref{onerankone}). (i) Local measurements (blue arrows) on
the physical sites $1$ to $ k_{1}$ generate the output quantum state
$ \vert \psi\rangle$ in the red virtual qubit at site $k_{1}+1$.
(ii) A filter measurement ${\cal F,\bar F}$ is applied at site
$k_{1}+1$. When not successful, local measurements (brown arrows) on
the physical sites $k_{1}+2$ to $k_{2}$ implement a unitary
transformation $V'= \vert 0\rangle\langle\varphi_{0} \vert + \vert
1\rangle\langle \varphi_{1} \vert $ on the correlation system, and
(ii) is repeated until the filtering operation is successful. (iii)
The unitary operation $V= \vert +\rangle\langle\varphi_{0} \vert +
\vert -\rangle\langle \varphi_{1} \vert $ is implemented by
measuring physical sites $k_{l} +2$ to $k'$. (iv) The filter
operation ${\cal F,\bar F}$ is now applied at site $k'_{1}+1$. When
not successful, local measurements (brown arrows) on the physical
sites $k'_{1}+2$ to $k'_{2}$ implement the unitary operation $V'$,
and (iv) is repeated until the filtering operation is successful. A
final measurement at site $k'_{l}+1$ in the basis $ \vert
m_{0}\rangle,  \vert m_{1}\rangle$ allows one to localize the
quantum information at site $k_{l}+1$.}\label{generalquwire}
\end{figure}

We now turn to the general universal 1D computational wires as
considered in \cite{Gr08}.  These are MPS having the canonical form
$A[0] \propto W$ and $A[1] \propto W\mbox{diag}(e^{-i\alpha},
e^{i\alpha})$, for some $W\in SU(2)$. For such MPS, there always
exists a basis $\left\{  \vert m_{s}\rangle \right\} $ such that
\begin{equation}
A[{m_{0}}] = r_{0} \vert \varphi _{0}\rangle \langle 0 \vert , \quad
A[{m_{1}}]  = r_{1} \vert \varphi _{0}\rangle \langle 0 \vert +
\vert \varphi_{1}\rangle \langle 1 \vert  \label{onerankone}
\end{equation}
where $r_{0}>0$, $r_1\geq 0$, and $r_{0} ^{2}+ r_{1} ^{2}=1$ (in
fact, these canonical forms are equivalent). Since we have a
universal computational wire, we can generate the desired output
state $ \vert \psi\rangle$ in the correlation system by suitable
local measurements on, say, the first $k$ sites, see
Fig.~\ref{generalquwire}. The remaining chain is in the state $\Phi(
\vert \psi\rangle)_{k+1}^N$.

(i) We can now rewrite
\begin{equation}
\Phi( \vert \psi\rangle)_{k+1}^N=\sum_{s=0,1}\lambda_s \vert
m'_{s}\rangle_{k+1} \Phi( \vert \varphi_s\rangle)_{k+2}^N
\end{equation}
where  $ \vert m'_{0}\rangle= r_{0} \vert m_{0}\rangle +r_{1} \vert
m_{1}\rangle$ and $ \vert  m'_{1}\rangle= \vert m_1\rangle$. This
situation is very similar as before (Eq. (\ref{phikplusone}));
however, we have to deal with non-orthogonal states as $\langle
m'_{0} \vert m'_{1}\rangle=r_{1}$. Let us now define a filter operation described by a local
two-outcome generalized measurement
$\{\mathcal{F},\bar{\mathcal{F}}\}$ with
\begin{align}
\mathcal{F} &=\tfrac{1}{\sqrt{1+r_1}}( \vert m_{0}\rangle \langle
m_{0} \vert +r_0 \vert m_{1}\rangle \langle m_{1} \vert  - r_1 \vert
m_1\rangle\langle m_0 \vert ) \\
\bar{\mathcal{F}} &= \sqrt{\tfrac{2 r_1}{1+r_1}} |\chi\rangle\langle \chi|,
\end{align}
where $ \vert \chi\rangle=\sqrt{(1-r_1)/2}
\vert m_0\rangle + \sqrt{(1+r_1)/2}  \vert m_1\rangle$ \cite{Wolfgang02}.
Note that $\bar{\mathcal{F}}$ has rank one. We now apply the filter
operation to site $k+1$. If the filter operation is successful, the
mapping $ \vert m_s'\rangle \to \sqrt{1-r_1}  \vert m_s\rangle$ is
accomplished (note that $1-r_1$ is the success probability). That
is, the basis states at site $k+1$ become orthogonal and the resulting
overall state of the chain is described by Eq. (\ref{phikplusone}). In case the filter
operation is not successful, the mapping $ \vert m_s'\rangle \to
\sqrt{r_1} \vert \chi\rangle$ is obtained, i.e. the measured site is
factored out. In this case, one uses part of the remaining wire to
implement the unitary operation $ \vert \varphi_s\rangle \to
 \vert s\rangle$ in the correlation system. The state of the remaining
wire is then again described by $\Phi( \vert \psi\rangle)_{k+1}^N$,
and we can repeat (i) until we succeed. The  success probability
after $l$ trials is given  $p= 1-r_{1}^{l}$. For any $\epsilon>0$,
one can achieve $p\geq 1-\epsilon$ with $l$ scaling efficiently as
$\log\frac{1}{\epsilon}$. Below we will see that $l$ is in fact
related to the correlation length of the resource state.

(ii) Without further touching site $k+1$, we use the remaining wire
(starting from $k+2$ to, say, site $k'$) to implement the unitary
operation  $ \vert \varphi_0\rangle \to  \vert +\rangle$, $ \vert
\varphi_1\rangle \to  \vert -\rangle$. After rewriting the state at
site $k'+1$ similarly as before, we obtain
\begin{equation}
\label{psik2} \sum_{s=0,1} Z_m^s \vert \psi_m\rangle_{k+1}
 \vert m'_s\rangle_{k'+1} \Phi( \vert \varphi_s\rangle)_{k'+2}^N,
\end{equation}

(iii) Again, the filter operation $\{\mathcal{F},\bar{\mathcal{F}}\}$
applied at site $k'+1$ allows one to obtain orthogonal states $
\vert m_s\rangle$, which can subsequently be distinguished, leading
in both cases to the state $ \vert \psi_m\rangle$ localized at site
$k+1$ (up to a Pauli correction). If the filter operation does not
succeed, site $k'+1$ is mapped to the state $ \vert \chi\rangle$ and
hence factored out. One may now use again part of the remaining wire
to implement the unitary operation $ \vert \varphi_s\rangle \to
 \vert s\rangle$ in the correlation system. The state of the remaining
wire is then again described by Eq. (\ref{psik2}), and we can repeat
(iii) until we succeed. The success probability after $l$ trials is
again given by $1-r_1^l$, and we have thus shown that one can indeed
efficiently localize the quantum information at a physical site.
\cite{footnote}

%\begin{figure}[tbh]
%\epsfig{file=Fig2.eps,width=9cm} \caption{ (Color online) Generation
%of arbitrary single-qubit output states from general quantum wires
%by Eq.~(\ref{onerankone}).}\label{generalquwire}
%\end{figure}

{\bf  Quantum computation with 2D resources.---} We now turn to 2D
resource states (computational webs) which are constructed by
coupling quantum wires vertically as in \cite{Gr08}. Each logical
qubit is associated with a particular (horizontal) wire, and the
computation proceeds by measuring sites from left to right, similar
to the one-way quantum computer. In a universal computational web,
it is possible to efficiently generate every $M$-qubit state $ \vert
\psi\rangle =U \vert 0\rangle^{\otimes M}$ resulting from a
poly-size quantum circuit $U$, in the correlation system. Using the
results established here for the 1D wires, we now argue that it is
always possible to further process the system such that the state $
\vert \psi\rangle$ is (efficiently) prepared on a subset of the
\emph{physical} sites. To do so, we again make use of the fact that
the remaining computational web is still universal. In fact, all we
need is that arbitrary single qubit rotations in the correlation
system can be achieved by processing individual wires. This allows
us to localize each qubit of $ \vert \psi\rangle$ to a physical site
by applying precisely the same procedure as discussed in the 1D case
to each wire independently. In order to do so, we make use of the
fact that the each wire carrying a logical qubit can be decoupled
from the remaining system \cite{Gr08}. In short, this shows that
every universal computational web state [12] is a universal state
preparator as well.

{\bf Extensions.---} %Although our protocols focus on the qubit
%resources of quantum computational webs \cite{Gr08}, extensions to
%more general resources --including all examples presented in \cite{Gr06,Gr07}-- are possible \cite{0902.1097}.
%%%%%%%%%%%%%%%%%%%%
%, including e.g. the AKLT state \cite{Gr07,Br08}. One can also extend the current
%protocol to prepare encoded physical states from the resources for
%encoded quantum computation in correlation space, where one logical
%qubit is encoded into several correlation systems, e.g. the modified
%toric code states \cite{Gr07}.
%%%%%%%%%%%%%%%%%%%%
%The common feature in these arguments seems to be that the very
%properties which must be present in a $\mathcal{QC}_{cs}$ in order
%to make the preparation of arbitrary states in the
%\emph{correlation} system possible, also make it possible to execute
%a post-processing protocol to eventually prepare output states in
%the \emph{physical} Hilbert space.
Although our protocols focus on the qubit resources of quantum
computational webs \cite{Gr08}, extensions to more general resources
--including all examples presented in \cite{Gr06,Gr07}-- are
possible. Consider e.g. 1D universal wires with $d \geq 2$ which
satisfy the following natural condition: there exists a basis
$\left\{  \vert e_{0}\rangle ,\cdots , \vert e_{d-1}\rangle \right\}
$ associated with each physical site such that the corresponding
operators $\{A[{e_{0}}], \cdots , A[{e_{d-1}}]\}$ are unitary up to
certain constant. Moreover, at least two operators $A[{e_{i}}]$ and
$A[{e_{j}}]$ are different in the sense that $A^{\dagger }[{e_{i}}]
A[{e_{j}}]\neq cI$, where $c$ is a constant. In this case one can
also localize quantum information to physical sites efficiently,
i.e. the corresponding $\mathcal{QC}_{cs}$ resources are in fact
universal state preparators. The only additional operation is the
local POVM $P= \vert m_{0}\rangle \langle m_{0} \vert + \vert
m_{1}\rangle \langle m_{1} \vert $, to cast the physical site into
the encoding subspace. An example of such a resource is the AKLT
state \cite{Gr07,Br08}. One can also extend the current protocol to
prepare encoded physical states from the resources for encoded
quantum computation in correlation space, where one logical qubit is
encoded into several correlation systems, e.g. the modified toric
code states \cite{Gr07}. The common feature in these arguments seems
to be that the very properties which must be present in a
$\mathcal{QC}_{cs}$ in order to make the preparation of arbitrary
states in the \emph{correlation} system possible, also make it
possible to execute a post-processing protocol to eventually prepare
output states in the \emph{physical} Hilbert space.

{\bf Entanglement criteria.---} Based on
the above protocols, one can conclude that many of the $\mathcal{QC}_{cs}$
universal states, in particular all examples presented in \cite{Gr06,Gr07,Gr08}, have the same quantum
information processing power as the universal resources in the
one-way model. That is, they allow one to efficiently reproduce the
\emph{quantum} output of each quantum computation by only local
operations, i.e., they are universal state preparators.
As a consequence, the entanglement-based criteria for (efficient
quasi-deterministic) universal state preparators are fully applicable
\cite{Va06a}. This imposes---contrary to what
was originally believed---rather severe entanglement requirements on
the universal $\mathcal{QC}_{cs}$ states. For example, the $\mathcal{QC}_{cs}$ universal resource states in
Refs.~\cite{Gr06,Gr07,Gr08}  must contain a high degree of entanglement in terms of either
Schmidt-rank width, geometric measure or Schmidt measure. All of these must
all scale faster than logarithmically with the system size---even though
e.g. the local entropy per site may be arbitrarily small and the two-point correlations may be non-vanishing.
Notice that in cases where output states are generated in an encoded form --as e.g. in the example of ``W-encoding'' in \cite{Gr06}--, the entanglement criteria for encoded universal state preparators \cite{Va06a} need to be applied.

{\bf  Role of correlation length.---} We have seen that for general
1D wires given by (\ref{onerankone}), a probabilistic
element---associated with the POVM $\{{\cal F}, \bar{\cal
F}\}$---enters in the protocol to prepare the output state
$|\psi\rangle$ in one of the physical sites. Next we illustrate that
this additional computational cost is related to the nonzero
correlation length $\xi$ of the resource state; the latter describes
qualitatively how far two distant physical sites still affect each
other. For nonzero correlation length, we have to ``concentrate''
information which has been distributed roughly over a distance $\xi$
during the computation.

As an example, consider an MPS resource with matrices $ A[0] =
 \cos\theta H $ and $ A[1] = \sin\theta H Z $ where $H$ is the Hadamard gate
 and $\theta \in ( 0,\frac{\pi}{4}]$; equivalently, there exists a basis $\{ \vert m_0\rangle ,  \vert
m_1\rangle \}$ such that $A[m_0] = \sin 2\theta  \vert +\rangle
\langle 0 \vert  , \; A[m_1] = \cos 2\theta  \vert +\rangle \langle
0 \vert  +  \vert -\rangle\langle 1 \vert $. Thus, this resource
falls into the more general class of Eq.~(\ref{onerankone}), with
$r_1 = \cos 2\theta$. Note that the 1D cluster state is obtained for
$\theta = \pi/4$. Following \cite{Fa92,Ve07}, we find that the
correlation length satisfies $e^{-1/\xi} = \sqrt{\cos 2\theta} =
\sqrt{r_{1}}$. The success probability to localize the output
quantum state after $l$ trials is $1-r_{1}^{l}$. In order to retain
the state preparation quasi-deterministic within the success
probability $1-\epsilon$, it is required that $l \geq
\frac{1}{2}\log \left(\frac{1}{\epsilon}\right) \xi$. In other
words, we have to keep the number $l$ of trials in proportion to the
correlation length $\xi$ (with a logarithmic factor dependent on
$\epsilon$), in order to cover distributed output information and to
convert it into a single physical site arbitrarily faithfully.

{\bf Conclusions.---} The broader context of this investigation is
to understand which features make a quantum computer more powerful
than a classical device. In the context of measurement based quantum
computation, this means to understand the features of universal
states that are necessary to guarantee their universality.
Interestingly, the $\mathcal{QC}_{cs}$ method showed that features
such as vanishing two-point correlations and a high degree of local
entanglement are \emph{not} necessary \cite{Gr06, Gr07, Gr08}.
However, the current results show that in many cases this method
cannot circumvent the stringent entanglement requirements from
\cite{Va06a}: we have shown that, under rather general conditions,
$\mathcal{QC}_{cs}$ universal resources are universal state
preparators, and hence must fulfill all entanglement requirements as
derived in \cite{Va06a}.

The results presented here cover the standard scenario of resources
where quantum information is processed in a 2D grid and individual
horizontal wires carry one logical qubit.
%However, it remains unclear whether similar arguments would also hold for resources
% where each logical qubit is more distributed over the lattice.
It remains unclear whether there exist computationally universal
resources where such a wire structure is no longer present but where
each logical qubit is highly de-localized, and if so, whether our
constructions can be extended to such a setting. More generally, it
would be interesting to understand whether resources for
measurement-based quantum computation exist that are computationally
universal, but do not allow to prepare the corresponding (encoded)
quantum output state, i.e. are not (encoded) universal state
preparators. Also the role of general local measurements as opposed
to projective measurements needs to be further investigated.

We acknowledge support from the FWF (J.-M. C. through the Lise
Meitner Program) and the European Union (QICS, SCALA). The research
at the Perimeter Institute is supported by the Government of Canada
through Industry Canada and by Ontario-MRI. MVDN acknowledges the
excellence cluster MAP.


\begin{thebibliography}{99}

\bibitem{Ra01}
R. Raussendorf and H.~J. Briegel, \emph{Phys. Rev. Lett.}
\textbf{86}, 5188--5191 (2001).

\bibitem{Br01}
H.~J. Briegel and R. Raussendorf, \emph{Phys. Rev. Lett.}
\textbf{86}, 910--913 (2001).

\bibitem{Br09}
H.J. Briegel et al., {\em Nature Physics} \textbf{51}, 19-26 (2009).

\bibitem{He06}
M. Hein et al., In \emph{Proceedings of the International School of
Physics ``Enrico Fermi'' on ``Quantum Computers, Algorithms and
Chaos''} (2005); arXiv:quant-ph/0602096.

\bibitem{Va06a}
M. Van~den Nest et al., \emph{Phys. Rev. Lett.} \textbf{97}, 150504
(2006); M. Van~den Nest et al., \emph{New J. Phys.} \textbf{9}, 204
(2007); C. E. Mora et al., arXiv: 0904.3641.

\bibitem{Ni05}
M. A. Nielsen, quant-ph/0504097; M. Van den Nest et al., \emph{Phys.
Rev. A} \textbf{77}, 012301 (2008).

\bibitem{Ba06}
S. D. Bartlett and T. Rudolph, \emph{Phys. Rev. A} \textbf{74},
040302(R) (2006).

\bibitem{Br08}
G. K. Brennen, A. Miyake, \emph{Phys. Rev. Lett.} {\bf 101}, 010502 (2008).

\bibitem{Chen08} X. Chen, B. Zeng, Z.-C. Gu, B. Yoshida, I.~L.
Chuang, arXiv:0812.4067.

\bibitem{Gr06}
D. Gross and J. Eisert, \emph{Phys. Rev. Lett.} \textbf{98}, 220503
(2007).

\bibitem{Gr07}
D. Gross et al., \emph{Phys. Rev. A} \textbf{76}, 052315 (2007).

\bibitem{Gr08}
D. Gross and J. Eisert, arXiv:0810.2542.

\bibitem{Fa92}
M. Fannes, B. Nachtergaele, and R. F. Werner, \emph{Comm. Math.
Phys.} \textbf{144}, 443--490 (1992).

\bibitem{Ve04}
F. Verstraete and J. I. Cirac, \emph{Phys. Rev. A} \textbf{70},
060302 (2004).

\bibitem{Ve07}
D. Perez-Garcia, F. Verstraete, M. M. Wolf, and J.I. Cirac,
\emph{Quant. Inf. Comp.} \textbf{7}, 401 (2007).

\bibitem{Wolfgang02} W. D\"{u}r, C. Simon, and J. I. Cirac, \emph{Phys. Rev. Lett.} {\bf 89}, 210402 (2002).

\bibitem{footnote}
We note that byproduct operators that occur in the
$\mathcal{QC}_{cs}$ scheme do not affect our protocol.

%\bibitem{0902.1097}  J.-M. Cai et al., arXiv: 0902.1097.
\end{thebibliography}
\end{document}